\documentclass[sigconf]{acmart}
\usepackage{booktabs} 
\usepackage{amsthm}
\usepackage{amsmath}
\usepackage{algorithm}
\usepackage{algorithmicx}
\usepackage{algpseudocode}

\hypersetup{draft}
\usepackage{textcomp}
\usepackage{graphicx}
\usepackage{subfigure}
\usepackage{caption}

\usepackage{multirow}
\usepackage{tabularx}
\usepackage{fancyhdr}
\usepackage{booktabs}

\copyrightyear{2021}
\acmYear{2021}
\setcopyright{acmcopyright}\acmConference[WSDM '21]{Proceedings of the
Fourteenth ACM International Conference on Web Search and Data Mining}{March
8--12, 2021}{Virtual Event, Israel}
\acmBooktitle{Proceedings of the Fourteenth ACM International Conference on Web
Search and Data Mining (WSDM '21), March 8--12, 2021, Virtual Event, Israel}
\acmPrice{15.00}
\acmDOI{10.1145/3437963.3441761}
\acmISBN{978-1-4503-8297-7/21/03}

\begin{document}

\title{Multi-Interactive Attention Network for Fine-grained Feature
 Learning in CTR Prediction}

{\author{Kai Zhang$^{1,3}$, Hao Qian$^3$, Qing Cui$^{3}$, Qi Liu$^{1,2}$}
\author{Longfei Li$^{3}$, Jun Zhou$^{3}$, Jianhui Ma$^{1}$, Enhong Chen$^{1,2,}$}}\authornote{Corresponding Author.}
\affiliation{$^{1}$Anhui Province Key Lab of Big Data Analysis and Application, School of Data Science, \\University of Science and Technology of China}

\affiliation{$^{2}$School of Computer Science and Technology, University of Science and Technology of China}
 \affiliation{$^{3}$Ant Group, Hangzhou, China}
\affiliation{kkzhang0808@mail.ustc.edu.cn, \{qiliuql, jianhui, cheneh\}@ustc.edu.cn\\
\{qianhao.qh, cuiqing.cq, longyao.llf, jun.zhoujun\}@antgroup.com}

\fancyhead{}

\begin{abstract}
In the Click-Through Rate (CTR) prediction scenario, user's sequential behaviors are well utilized to capture the user interest in the recent literature. However, despite being extensively studied, these sequential methods still suffer from three limitations. First, existing methods mostly utilize attention on the behavior of users, which is not always suitable for CTR prediction, because users often click on new products that are irrelevant to any historical behaviors. Second, in the real scenario, there are numerous users that have operations a long time ago, but turn relatively inactive in recent times. Thus, it is hard to precisely capture user's current preferences through early behaviors. Third, multiple representations of user's historical behaviors in different feature subspaces are largely ignored.

To remedy these issues, we propose a \emph{Multi-Interactive Attention Network} (MIAN) to comprehensively extract the latent relationship among all kinds of fine-grained features (e.g., gender, age and occupation in user-profile). Specifically, MIAN contains a \emph{Multi-Interactive Layer} (MIL) that integrates three local interaction modules to capture multiple representations of user preference through sequential behaviors and simultaneously utilize the fine-grained user-specific as well as context information. In addition, we design a \emph{Global Interaction Module} (GIM) to learn the high-order interactions and balance the different impacts of multiple features. Finally, Offline experiment results from three datasets, together with an Online A/B test in a large-scale recommendation system, demonstrate the effectiveness of our proposed approach. 

\end{abstract}



\maketitle
{\fontsize{8pt}{8pt} \selectfont
	\textbf{ACM Reference Format:}\\
	Kai Zhang, Hao Qian, Qing Cui, Qi Liu, Longfei Li, Jun Zhou, Jianhui Ma, Enhong Chen. 2020. Multi-Interactive Attention Network for Fine-grained Feature Learning in CTR Prediction. In \textit{Proceedings of the Fourteenth ACM International Conference on Web Search and Data Mining (WSDM'21), March 8–12, 2021, Virtual Event, Israel}. ACM, New York, NY, USA, 10 pages. \\ https://doi.org/10.1145/3437963.3441761
}

\section{Introduction} 
CTR prediction, which aims to estimate the likelihood of a user clicking at an ad or an item, is critical to many applications such as online advertising \cite{mcmahan2013ad,he2014practical} and recommendation systems \cite{9312492,liu2012cocktail,liu2011personalized}. In the recommender systems, the light improvement of the CTR prediction has a huge impact on the revenue of the business. Thus, it has drawn more and more research attention from the communities of academia and industry to develop efficient algorithms.

\begin{figure}
        \vspace{0.25cm}
	\centering
	\includegraphics[width=0.85\columnwidth]{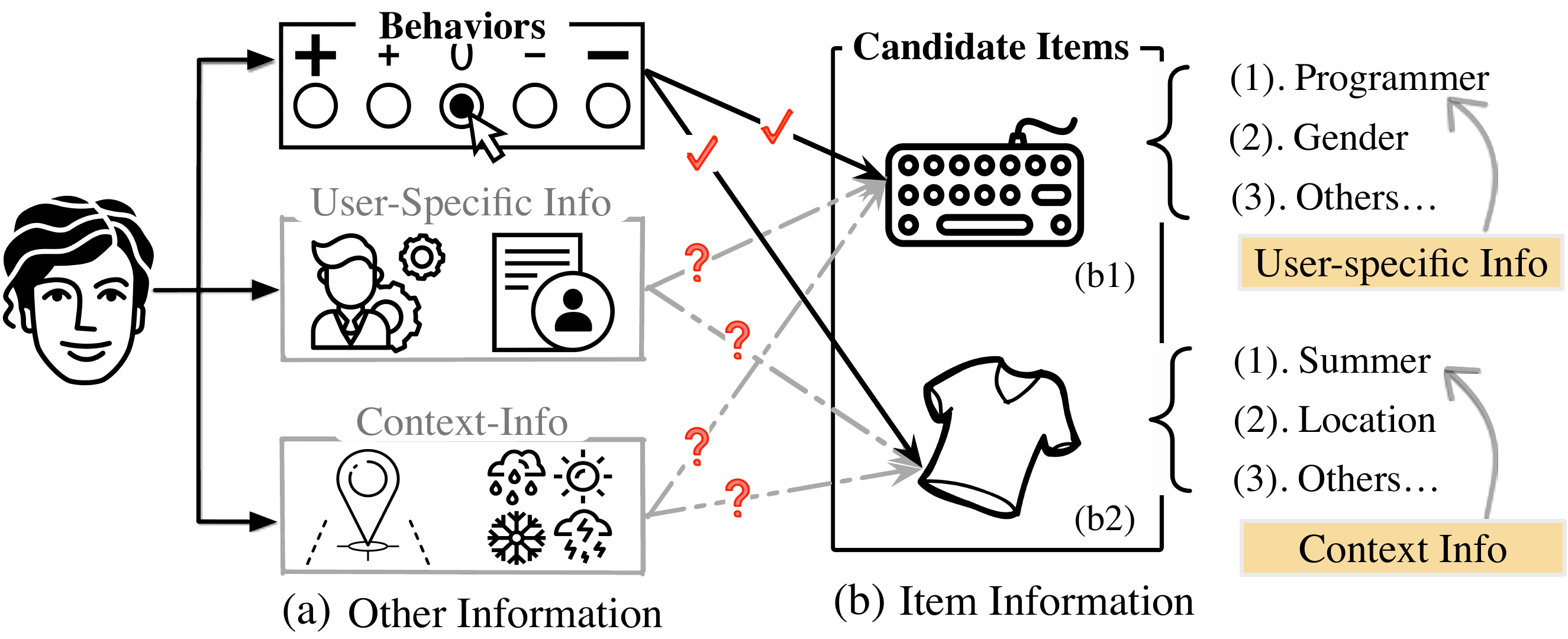}
	\vspace{-0.2cm}
	\caption{\small{Illustration of the associations (i.e., lines between (a) and (b), the check mark represents the relationship that has been used, and the question mark represents not been fully explored) between candidate items and other fine-grained information. 
	}}
	\label{fig:example}
	\vspace{-0.45cm}
\end{figure}

In the literature, there are numerous efforts for this problem  \cite{chen2017locally,huang2017ad,he2017neural}. Among them, supervised machine learning methods are proposed to learn shallow feature combinations, such as \emph{Logistic Regression} (LR), \emph{Factorization Machine} (FM) \cite{rendle2010factorization} and its variants \emph{Field-aware FM} (FFM) \cite{juan2016field}.
In the last few years, some deep learning models have been proposed to learn sophisticated feature interactions in CTR predictions. For example, \emph{Wide \& Deep} \cite{cheng2016wide} leverages feature engineering in the wide part (i.e., LR) while generating high-order features through deep neural networks. \emph{Deep \& Cross Network} (DCN) \cite{wang2017deep} and DeepFM \cite{guo2017deepfm} extends it by replacing the wide part with more practical and robust networks. 
Recently, some sequential methods utilize the \emph{Recurrent Neural Network}-based attention to extract the representation of user interest through the historical behaviors, such as \emph{Deep Interest Net} (DIN) \cite{zhou2018deep} and \emph{Deep Interest Evolution Net} (DIEN) \cite{zhou2019deep}. 
In summary, the high-order interest excavate of user's historical behaviors significantly enhances the representation capability of the models and further improves the performance of the CTR prediction.

However, as shown in Figure~\ref{fig:example}, the sequential methods above mainly focus on mining the associations between candidate item and the historical behaviors, but pay little attention to some equally valuable fine-grained user-specific (e.g., age, gender, and occupations) and context (e.g., weather and location) information, which may suffer from some limitations:
\begin{itemize}
	\item Most previous approaches explore user's interests from one's historical behavior, which may deviate from the real preference and mislead the CTR prediction because users often have new demands that are irrelevant to any historical behaviors in the e-commerce systems (e.g., Figure~\ref{fig:data} (a)).
	\item In the real CTR scenario, many user's click behaviors mostly happen a long time ago and lack of activities in recent periods (e.g., Figure \ref{fig:data} (b)). Thus, solely relying on the historical behaviors of users may lead to outdated recommendations that do not go with the user's current demands.
	\item Finally, as for sequential behavior encoding, the diversity of feature interactions in different subspaces has not been fully considered, i.e., the representation of historical behaviors in different click-through contexts should be different. For example, the ``T-shirt'' may be activated as a user's behavior representation in ``summer'', rather than ``winter''.
\end{itemize}

Fortunately, there exists a large amount of user-specific and context information in many recommender systems. This fine-grained information provides various clues to infer the user's current state which can bring a significant improvement of performance to personalized recommendations \cite{rendle2011fast,shi2014cars2}, especially when the historical behavior is limited or unrepresentative. 
For instance, in Figure 1 (b1), the candidate item ``mechanical keyboard'' may be more relevant to users' current occupation ``programmer'' which is hard to represent in the historical behaviors. Therefore, learning the fine-grained feature interactions helps model the current state of preference of users, thus can largely eliminate the above drawbacks of the previous sequential methods in CTR prediction. 

In this paper, we propose a novel model named \emph{\textbf{M}ulti-\textbf{I}nteractive \textbf{A}ttention \textbf{N}etwork} (MIAN), which can aggregate multiple information, and gain latent representations through interactions between candidate items and other fine-grained features. 
To be specific, the network consists of a \emph{Multi-Interactive Layer} (MIL) which includes three local interaction modules and a global interaction module. The first local module is \emph{Item-Behaviors Interaction Module} (IBIM) that uses Pre-LN Transformer to adaptively explore the user preferences of sequential behaviors at different subspaces. The second is \emph{Item-User Interaction Module} (IUIM) which aims to capture the knowledge between candidate items and the fine-grained user-specific information. Similarly, the third named \emph{Item-Context Interaction Module} (ICIM) is devised to mine relations between candidate items and context-aware information. Besides, the \emph{Global Interaction Module} (GIM) is designed to study and weigh the influence between the low-order features after the embedding layer and high-order features generated from three local interaction modules. To summarize, the main contributions can be highlighted as follows: 
\begin{itemize}
	\item We propose to study the problem of multiple fine-grained feature interactions as well as user's historical behaviors simultaneously, that, to our knowledge, has not been explicitly and jointly modeled by previous CTR methods.
	\item We design a novel MIAN model, which contains a multi-interactive layer for fine-grained feature interactive learning, and a Transformer-based module to extract multiple representations of user behavior in different feature subspaces.
	\item Extensive experiments on three large datasets demonstrate that our approach not only outperforms the state-of-the-art CTR methods significantly, but also has good model explainability. Moreover, we deploy MIAN on a large commercial system and achieves significant improvement.
\end{itemize}

\section{related work} 
In the early stage of recommendation systems, people spent much time on tedious and onerous feature engineering. At that time, the dimensions of the raw features are relatively small, which makes it possible to implement different combinations of raw features. The newly created features are then fed into a shallow model, such as \emph{Logistic Regression} (LR) \cite{rendle2010factorization} and \emph{Gradient Boosting Decision Trees} (GBDT) are widely used in CTR prediction. Then, \emph{Factorization Machine} (FM) \cite{rendle2010factorization} transforms the learning of users and items features into a small, shared vector shape. Based on the above method, \emph{Field-aware} \cite{juan2016field} and \emph{Field-weighted FM} (FFM) \cite{pan2018field} further consider the different impact of the fields that a feature belongs to in order to improve the performance of CTR prediction. Along this line, \emph{Attentional Factorization Machines} (AFM) \cite{xiao2017attentional} are proposed to automatically learn weights of deep cross-features and \emph{Neural Factorization Machines} (NFM) \cite{he2017neural} enhances FMs by modelling higher-order and non-linear feature interactions. 

\begin{figure*}
	\vspace{0.15cm}
	\centering
	\includegraphics[width=2\columnwidth]{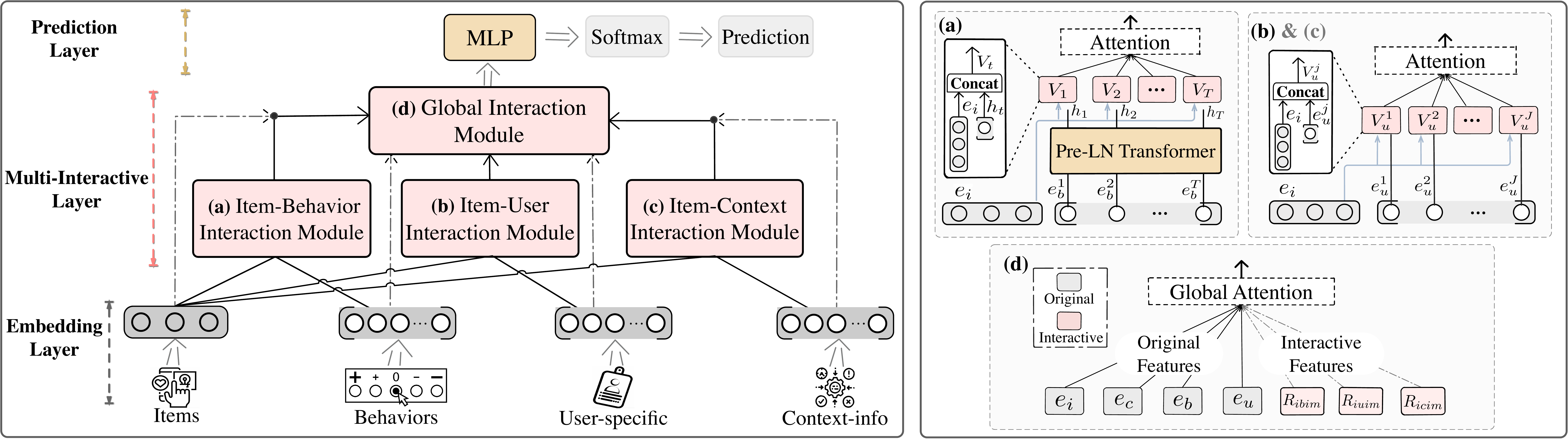}
	\caption{\small{The architecture of MIAN. Overall, from the bottom up, it can be divided into three layers: 1. Embedding Layer, which projects sparse heterogeneous features into low-dimensional vectors. 2. Multi-Interactive Layer, which contains three local modules, i.e., (a), (b), (c), and a global module, i.e., (d), to learn multiple fine-grained feature interactions. 3. Prediction Layer, which contains DNN to do CTR prediction.
	}}
	\label{fig:model}
\end{figure*}

Recently, the success of deep neural networks (DNN) in natural language processing \cite{zhang2019interactive,tao2019radical,jin2019promotion} and computer vision \cite{krizhevsky2012imagenet,ciresan2012deep,rastegari2016xnor} brings a new direction to the recommendation system. Among them, the \emph{Wide \& Deep learning} \cite{cheng2016wide} introduces deep neural networks to the CTR prediction. It jointly trains a deep neural network along with the traditional wide linear model. The deep neural networks liberated people from feature engineering while generalizing better combinations of the features. Lots of variants of the \emph{Wide \& Deep learning} have been proposed since it revolutionizes the development of the CTR prediction. \emph{Deep \& Cross network} (DCN) \cite{wang2017deep} replaces the wide linear part with cross-network, which generates explicit feature crossing among low and high level layers. DeepFM \cite{guo2017deepfm} combines the power of DNN and factorization machines for feature representation in the recommendation system. Furthermore, xDeepFM \cite{lian2018xdeepfm} extends the work of DNN by proposing a Compressed Interaction Network to enumerate and compress all feature interactions. Overall, the deep models mentioned above all construct a similar model structure by combing the low-order and high-order features, which greatly reduce the effort of feature engineering and improve the performance of CTR prediction. 

However, these aforementioned shallow or deep models take statistical and categorical features as input, while discarding the sequential behavior information of users. For example, users may search items at an e-commerce system, then view some items of interest, and these items are likely to be clicked or purchased next time. 
Since the historical behaviors explicitly indicate the preference of the users, it has gained much more attention in the recommendation systems \cite{li2018learning,pan2016sparse,zhou2018deep,zhou2019deep}. Among them, DIN \cite{zhou2018deep} proposes a local activation unit that learns the dynamic user interests from the sequential behavior features. DIEN \cite{zhou2019deep} designs an interest evolving layer an attentional update gate to model the dependency between sequential behaviors.
The researches above realized the importance of user's historical behaviors. Unfortunately, they just project other information (i.e., user-specific and context) into one vector and did not pay equal attention to the interactions between the candidate item and fine-grained information, while modeling this interaction has shown extensively progress in many tasks, such as search recommendations \cite{biancalana2011enhancing,gasparetti2017personalization} and knowledge distillation \cite{chen2017learning,chen2017spatial}.

Different from all previous methods, our MIAN can explore the sequential behavior and other fine-grained information simultaneously. Specifically, compared to shallow and deep models \cite{rendle2010factorization,juan2016field,wang2017deep,guo2017deepfm,lian2018xdeepfm}, MIAN has a remarkable ability to encode user's preference through sequential behavior. Compared to sequential models \cite{zhou2018deep,zhou2019deep}, MIAN can model fine-grained feature interactions better when historical behavior is not enough or representative.

\section{The MIAN Model} 

\subsection{Problem Definition} 
\label{section:problem_define}
First, we define the problem of the CTR prediction. It estimates the probability that a user clicks at candidate items based on the input feature representations. A sequential CTR model mostly takes four groups of features as input: $CTR = f(candidate\ item, behaviors, use\\r\ profile, context)$. Different from previous methods, we utilize several fine-grained features in groups. For instance, user-specific fields includes gender, age, occupation, and season, etc. Context fields consist of log time, location along with some other features. Formally, in our model, to model the candidate item interacting with other fine-grained features with attention mechanism, the candidate item information is concatenated to one field represented by $x_i$. Moreover, user-specific information is denoted as $x_u = [ x_u^1,.., x_u^j, ..., x_u^J]$ with $J$ fine-grained fields in total. Context information is denoted as $x_c = [ x_c^1,..., x_c^k, ..., x_c^K]$ with $K$ fine-grained fields, and user behavior is represented by $x_b = [ x_b^1,.., x_b^t, ..., x_b^T]$ of $T$ consecutive sequential behaviors. The goal of CTR prediction is to predict the probability $\hat{y}$ that user $u$ clicks at the candidate item $i$, given the above four groups of features.

\subsection{Components of MIAN}
\label{section:components}
Figure~\ref{fig:model} illustrates the overall architecture of MIAN, which consists of three layers: a commonly used Embedding Layer, a novel Multi-Interactive Layer, and a Prediction Layer. The details of each layer will be explained in the following sections. 

\subsubsection{\textbf{Embedding Layer.}}
Most data from industrial recommendation systems are presented in a multi-group form. Hereby, like most sequential models \cite{zhou2018deep,zhou2019deep}, we divide the raw data into four groups: candidate item, historical behavior, user-specific, and context information. Initially, the raw data are represented as sparse vectors, which contain the information of multiple fine-grained fields mentioned in section~\ref{section:problem_define}. Specifically,
\begin{equation}
	\begin{split}
	& x_* = [x_*^1;...;x_*^i; ...; x_*^S],
	\end{split}
	\label{eq:three}
\end{equation}
assume $S$ is the number of total fields in each group, and $x_*^i$ is the feature representation of the $i$-th field. Note that, $x_*^i$ is a one-hot vector if the $i$-th field is categorical. $x_*^i$ is a set of scalar values if the $i$-th field is numerical. Since the feature representations of the categorical features are sparse and high-dimensional, they are projected into low-dimensional spaces, which can be formulated as follows:
\begin{equation}
	e_*^i =  
	\left\{
             \begin{array}{lr}
             {V_i}{x_*^i} ,  \ \  {x_*^i} \ \ \text{is one-hot vector}, &\\
              \\
             {x_*^i},  \ \  {x_*^i} \ \ \text{is scalar vector}, 
             \end{array}
	\right.
	\label{eq:four}
\end{equation}
where $V_i$ is an embedding matrix and $e_*^i$ is the embedding vector of fine-grained field $i$. By doing this, the output of the embedding layer would be a concatenation of multiple embedding vectors, which can be represented as $e_i$, $e_b = [e_b^1;...;e_b^t; ...; e_b^T]$, $e_u = [e_u^1;...;e_u^j; ...; e_u^J]$, $e_c = [e_c^1;...;e_c^k; ...; e_c^K]$.

\subsubsection{\textbf{Multi-Interactive Layer.}} 
In this section, we will explain the MIL which models the relations between the candidate item and the other information in an efficient way. To be specific, the IBIM captures the evolution of user's interest preference through the sequential behavior. Furthermore, the fine-grained features in user-specific and context information can generate more features through interactions with the candidate item in IUIM and ICIM. Next, we describe these local modules in detail.

\textbf{Item-Behaviors Interaction Module (IBIM)}. In CTR prediction literature with regrading to sequential behavior, it has shown great potential to model the interaction between the candidate item and user's sequential behavior \cite{zhou2018deep,zhou2019deep}. Therefore, we propose the \emph{Item-Behaviors Interaction Module} which contains a modified Transformer unit and an attention mechanism to model these two types of information simultaneously. Note that despite the strong capability of Transformer in feature learning, the model efficiency and complexity are retained according to Vaswani et al. \cite{vaswani2017attention}. Moreover, we only use one Transformer block in IBIM module, and the increased amount of parameters than previous RNN-based sequential models (e.g., DIN and DIEN) is relatively small compared to the parameter size of the embedding layer. 

As shown in Figure~\ref{fig:model} (a), we employ the Pre-LN Transformer which shows faster and stabler convergence than the original Transformer layer in theoretical research~\cite{wang2019learning,xiong2020layer} and our experiments. The Pre-LN Transformer unit is slightly different from the one introduced in BERT \cite{devlin2018bert}. To be specific, by stacking the layer normalization, multi-head self-attention, and position-wise feed-forward networks together in a certain order, it becomes the well-known Transformer unit. In contrast with the structure of the original Transformer, the layer normalization is applied before feeding inputs to multi-head self-attention and position-wise feed-forward network. The computational processes are detailed as follows:

\vspace{0.1cm}
\emph{1) Layer Normalization}. We first normalize the input by feeding it to the layer normalization network for accelerating and stabilizing the training process of the whole network. The layer normalization transformation is defined as:
\begin{equation}
    \centering
    l_b = \gamma \odot \frac{e_b - \mu}{\sqrt{\sigma^2 + \epsilon}} + \beta\ ,
    \label{eq:five}  
\end{equation}
where $\mu$ and $\sigma$ are the mean and variance of the input matrix $e_b$, $\odot$ is the element-wise multiplication. $\gamma$ is a learnable scale factor and $\beta$ is the learnable bias term. $\epsilon$ is a small value added as the smooth factor to avoid dividing by zero. The output of the layer normalization network is denoted as $l_b$, which is then fed to the multi-head self-attention network.

\emph{2) Multi-Head Self-Attention.}
The multi-head self-attention mechanism has received a wide range of applications in natural language processing \cite{devlin2018bert}. Multi-head self-attention is capable of learning from multiple subspace representations at different time steps jointly. We adopt this attention mechanism with $h$ heads in this work. The product from each self-attention network is then concatenated and linearly transformed as follows:
\begin{equation}
    \centering
    \begin{split}
	& MultiHead(x) = Concat(head_1, head_2,..., head_h) W^O ,\\
    & head_i = Attention(l_b W^Q_i, l_b W^K_i, l_b W^V_i)\ ,
    \end{split}
	\label{eq:six}
\end{equation}
where $i$ denotes the $i$-th self-attention subspace, $W^Q_i$, $W^K_i$, $W^V_i$, and $W^O$ are learnable projection parameters. The output of the multi-head self-attention network is represented by $m_b$. 

\vspace{0.1cm}
\emph{3) Position-wise Feed-Forward Network (FFN).}
Since the multi-head self-attention network is a series of linear transformations, the position-wise feed-forward network then applies the non-linear transformations on the output of the multi-head self-attention network. The FFN consists of two linear transformations along with a ReLU activation in between. More formally:
\begin{equation}
    \centering
    FFN(m_b) = max(0, m_b W_1 + b_1)W_2 + b_2\ ,
    \label{eq:seven}
\end{equation}
where $W_1$, $b_1$, $W_2$ and $b_2$ are learnable parameters in the linear transformations. Until now, with the input $e_b$, we can get the output $h_b$ (i.e., $[h_1; ...; h_t; ...; h_T]$) of the Pre-LN Transformer unit. 

For the hidden states of user behaviors from the Pre-LN Transformer unit, we simply concatenate them with the candidate item representation $e_i$ to get the interactive feature representation at each time step, formulated by:
\begin{equation}
	\begin{split}
	& V_t = Concat(e_i,h_t).
	\end{split}
	\label{eq:eight}
\end{equation}

As we discussed before, user interest varies over time so that each historical behavior may leave a different influence on the representation of the user interest. Therefore, it is necessary to quantify the contribution of each user behavior. Fortunately, the attention mechanism emphasizes different parts of user behaviors by assigning weights to the behavior representation $V_t$. We can generate the attention weight $\alpha_t$ of each behavior as follows: 
\begin{equation}  
	\begin{split}
	&\alpha_t=\frac{\exp(\gamma (V_t))}{\sum_{t=1}^{T} \exp(\gamma (V_t))}\ ,
	\end{split}
	\label{eq:nine}
\end{equation}
\begin{equation}
	\begin{split}
	&\gamma(V_t)={\tanh}({V_t}\ \cdot\ W_t\ +\ {\widehat{b}_t})\ ,
	\end{split}
	\label{eq:ten}
\end{equation}
where $\gamma$ is score function. $W_t$ and $\widehat{b}_t$ are weight and bias matrix respectively. $tanh$ is a non-linear activation function. In this way, the attention weight $\alpha_t$ greatly enhances the explanatory ability of MIAN. It enables us to extract behaviors with high attention weight, which is helpful for user behavior modelling. After computing the attention weights, we can further get the user interest representation through the interaction with the candidate item. And the final output feature representation is formulated as:
\begin{equation}
	\begin{split}
	&R_{ibim}=\sum_{t=1}^{T}{\alpha_t}{V_t}\ .
	\end{split}
	\label{eq:eleven}
\end{equation}

\textbf{Item-User Interaction Module (IUIM)}. In the real CTR prediction scenario, the candidate item may be associated with the user-specific information. Since the user's preference may be missing in the user sequential behaviors, it is necessary to compensate for this information through user-specific information. However, most previous sequential methods only focus on mining the historical behaviors, which lead to one common limitation, i.e., they do not perform very well in the absence of sequential data. To remedy the issue, we add the IUIM (Figure~\ref{fig:model} (b)) to further mine the interactions between candidate items and fine-grained user-specific information. Specifically, we concatenate the item representation $e_i$ and fine-grained user-specific representations of each field $\{e_u^1, ..., e_u^j, ..., e_u^J\}$ to get $\{V_u^1, ..., V_u^j, ..., V_u^J\}$, where $V_u^j = Concat(e_i, e_u^j)$. 

Next, we quantify the impact of user-specific information on candidate item information by generating the attention weight $\alpha_j$ with the following formula: 
\begin{equation}  
	\begin{split}
	&\alpha_j=\frac{\exp({\tanh}({V_u^j}\cdot W_u^j\ +\ {\widehat{b}_u^j}))}{\sum_{j=1}^{J}\exp({\tanh}({V_u^j}\cdot W_u^j\ +\ {\widehat{b}_u^j}))}\ ,
	\end{split}
	\label{eq:twelve}
\end{equation}
where $W_u^j$ and $\widehat{b}_u^j$ are weight and bias matrix respectively. Through the attention weight $\alpha_j$, the fine-grained user-specific information that best match the candidate item can be learned and further extracted. With the attention weight $\alpha_j$, the final interactive representation between the candidate item and user-specific information is formulated as:
\begin{equation}
	\begin{split}
	&R_{iuim}=\sum_{j=1}^{J}{\alpha_j}{V_u^j}\ .
	\end{split}
	\label{eq:thirteen}
\end{equation}

\textbf{Item-Context Interaction Module (ICIM)}. As we mentioned before, most researches in CTR prediction task pays little attention on the interaction between candidate items and fine-grained context information. However, the fine-grained features in context information, such as weather and season, are closely associated with the activeness of candidate item. For instance, the sales of {``t-shirts''} surge in the {``summer''}, while diminish in {``winter''}. Thus, it is valuable to incorporate the interactions between candidate items and the fine-grained context information for the CTR prediction. Similar to IUIM architecture, we introduce ICIM to quantify the importance of this factor and calculate the scaled attention representation as follows:
\begin{equation}
	\begin{split}
	& V_c^k = Concat(e_i, e_c^k) ,
	\end{split}
	\label{eq:eight}
\end{equation}
\begin{equation}
	\begin{split}
	&R_{icim}=\sum_{k=1}^{K}{\ \frac{\exp({\tanh}({V_c^k}\cdot W_c^k\ +\ {\widehat{b}_c^k}))}{\sum_{k=1}^{K}\exp({\tanh}({V_c^k}\cdot W_c^k\ +\ {\widehat{b}_c^k}))}\ }{V_c^k}\ ,
	\end{split}
	\label{eq:fourteen}
\end{equation}
where $W_c^k$ and $\widehat{b}_c^k$ are weight and bias matrix. $V_c^k$ is the concatenation of each fine-grained field of context information $e_c^k$ with the candidate item representation $e_i$. The $R_{icim}$ represents the feature interaction between the candidate item and context information. 

\textbf{Global Interaction Module (GIM)}. At this stage, we obtain both the original feature representation (i.e., $e_*$) from the Embedding Layer and the generated interactive feature representation (i.e., $R_*$) from three local interaction modules. As the effectiveness of Deep\&Cross \cite{wang2017deep} network indicates the importance of the interactions between low-order and high-order features. To inherit this idea, as shown in Figure~\ref{fig:model} (d), the Global Interaction Module is designed to explicitly capture the relationship between the original feature (i.e., low-order) and the generated interaction feature (i.e., high-order). Moreover, we utilize the attention mechanism to extract the importance of different interaction modules and original embedding. To be specific, we first concatenate all features into one representation vector $r_{g}$, which can be formulated as:
\begin{equation}
	\begin{split}
	& r_g = \{e_i;\ e_c;\ e_b;\ e_u;\ P(R_{ibim});\ P(R_{iuim});\ P(R_{icim})\} \\
	& \ \ \ \ = [r_1;\ r_2;\ r_3;\ r_4;\ r_5;\ r_6;\ r_7],
	\end{split}
	\label{eq:fifteen}
\end{equation}
where $P(\cdot)$ is the pooling operation. The combinations of all the features are then taken as the inputs of GIM, which utilizes the global attention unit to extract the relationship among different parts of the input feature. The output is calculated by:
\begin{equation}
	\begin{split}
        &R_g=\sum_{l=1}^{L}{\ \frac{\exp({\tanh}({r_l}\cdot W_l\ +\ {\widehat{b}_l}))}{\sum_{l=1}^{L}\exp({\tanh}({r_{l}}\cdot W_{l}\ +\ {\widehat{b}_{l}}))}\ }{r_l}\ ,
	\end{split}
	\label{eq:sixteen}
\end{equation}
where $W_l$ and $\widehat{b}_l$ are weight and bias matrix respectively. $r_l$ is the representation of each individual feature, $L$ is the number of total features in $r_g$. $R_g$ is the global representation of all features.

\subsubsection{\textbf{Prediction Layer.}} 
In the end, we design a prediction layer which contains a DNN prediction module to further enhances the capability of mining the high-order feature.

\vspace{0.1cm}
\textbf{DNN Prediction Module}. In this module, the generated global feature representation $R_g$ is fed to a DNN prediction module, which utilizes a commonly used Multilayer Perceptron (MLP) for the final CTR prediction. Specifically, we stack multiple fully connected layers with non-linear activation function in between to automatically learn the high-order interactions among the input features. The process is formulated as:

\begin{equation}
R_{l}=\text{Relu} ({W}_{l} {R_{l-1}}+b_{l}), l=1,2,\cdots,h
\label{eq:senventeen}
\end{equation}
where $W_i, b_i$ are trainable parameter matrices and biases in the hidden layers and $R_h$ is the output of $h$-th layer which is also regard as the output of the DNN prediction module. For the last hidden layer, we will make final CTR predictions (i.e., $\hat{y}$) through $softmax$ function, which can be represented as follows: 
\begin{equation}  
	\begin{split}
	&\hat{y} = softmax(W_q R_h + b_q).
	\end{split}
	\label{eq:eighteen}
\end{equation}

At this stage, our proposed MIAN generates the final CTR prediction after performing feature interactions through the Embedding, Multi-Interactive, and Prediction Layer on the original four feature groups including candidate item, user sequential behaviors, user-specific and context information.

\subsection{Training Strategy}
To estimate parameters of a model, we need to specify an objective function to optimize. For better comparison, we specify a traditional loss function which is widely used in previous works \cite{guo2017deepfm,lian2018xdeepfm,zhou2018deep,zhou2019deep} for model training. The goal of the objective function is to minimize the cross-entropy of the predicted values and the real labels, which is defined as:
\begin{equation}
	\begin{split}
	&\mathcal{L}(y, \hat{y})=- y log\hat{y} - (1-y)log(1-\hat{y})\ ,
	\end{split}
	\label{eq:nineteen}
\end{equation}
where $y \in \{0, 1\}$ is the ground truth and  $\hat{y} \in (0, 1)$ is the predicted probability of y. Additionally, all the parameters are optimized by the standard back-propagation algorithm \cite{lecun2015deep}.

\section{Experiments} 
In this section, we present the details of the experiments, including datasets description, baseline methods, experiment setups and the analysis of multiple experiment results.

\subsection{Datasets Description}
Experiments are conducted on two public datasets from Amazon and a commercial dataset collected from an industrial recommendation system as well. We discard data of users and items with very few historical behaviors to avoid data sparsity. The basic statistics of all the datasets are shown in table~\ref{tab:data}. 

\subsubsection{\textbf{Amazon Dataset}} (McAuley et al. 2015) is composed of product reviews and metadata from Amazon. We use two subsets of Amazon dataset, with abundant sequential behaviors, user-specific and context information, to study the performance of MIAN. In these datasets, the reviews of users sorted by the log time are taken as the user sequential behaviors. 
As there are $T$ behaviors of user $u$ in total along with multiple user-specific and context attributes. We take the first $T-1$ behaviors as the input features in order to predict whether a user will make a $T$-th behavior by writing reviews.

\begin{table}
    \small
    \centering
    \vspace{0.1cm}
    \caption{Statistics of datasets used in this paper.} 
    \vspace{-0.1cm}
    \label{tab:data} 
    \begin{tabular}{lrrrrp{1.8cm}p{1.2cm}p{1.2cm}p{1.2cm}p{1.2cm}}
            \toprule
            \specialrule{0em}{2pt}{0pt}
            \multirow{2}{*}{\large{Dataset}} &\multicolumn{4}{c}{Data Attributes}\\ 
           \specialrule{0em}{1pt}{0pt}
            \cline{2-5}\specialrule{0em}{3pt}{0pt}&\ \ \#\ User&\ \#\ Item.&\ \#\ Cate.&\ \#\ Samp.\\
            \midrule
            Amazon (Book). &\ \ 603,668&\ \ 367,982&\ \ 1,600&\ \ 8,900,038\\
            \specialrule{0em}{3pt}{0pt}
            Amazon (Electro). & \ \ 192,403&\ \ 63,001&\ \ 801&\ \ 1,689,188\\
            \specialrule{0em}{3pt}{0pt}
            Commercial. & \ \ 2,163,147 &\ \ 41 &\ \ 41 &\ 36,096,332\\
            \bottomrule
    \end{tabular}
\end{table}

\begin{figure}
	\centering
	\includegraphics[width=1\columnwidth]{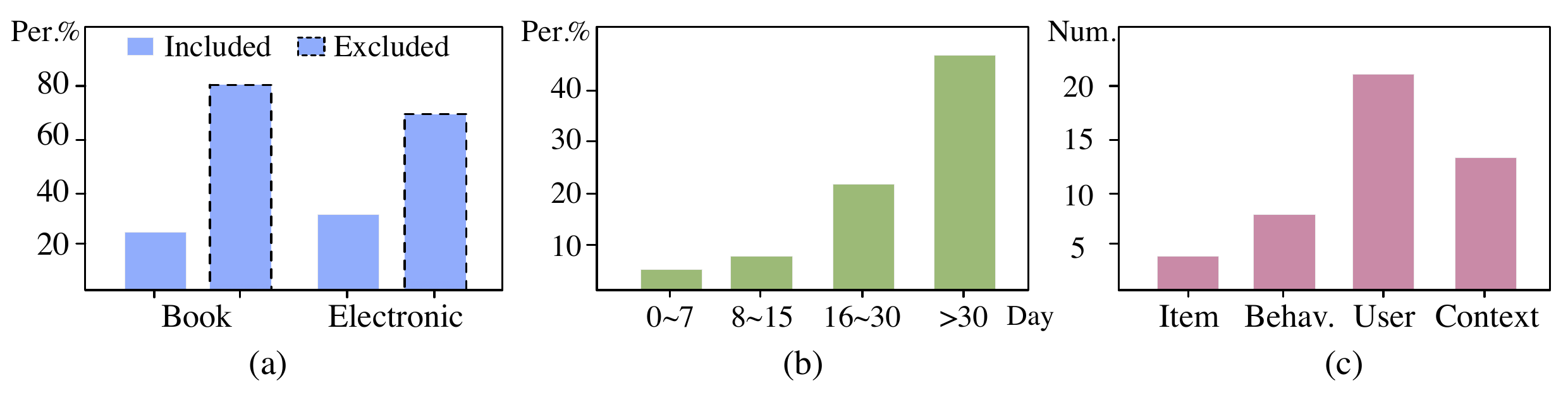}
	\vspace{-0.55cm}
	\caption{\small{{(a)} The probability that the candidate item is included in the previous behavior. {(b)} In Amazon datasets, number distributions of user's historical behavior. {(c)} The number of fine-grained attributes in different feature groups in the commercial dataset.}}
	\label{fig:data}	
	\vspace{-0.3cm}
\end{figure}

\subsubsection{\textbf{Commercial Dataset}} is collected from click logs from the Alipay recommendation system\footnote{\ \ \ \
The data set does not contain any Personal Identifiable Information (PII).

The data set is desensitized and encrypted.

Adequate data protection was carried out during the experiment to prevent the risk of data copy leakage, and the data set was destroyed after the experiment. 

The data set is only used for academic research, it does not represent any real business situation.
}. We sampled the items that users viewed in a time period of 30 days as the training samples. A sample instance consists of the candidate item, user-specific, context and its historical behaviors. The items that user clicked in the previous 29 days contribute to user sequential behaviors. The goal is to predict whether the user clicks at the candidate item on the last day. 

\vspace{0.1cm}
Figure~\ref{fig:data} illustrates number distributions of the candidate item, historical behavior and fine-grained attributes, respectively. We can observe some phenomenons in the real CTR scene: Specifically, first, there are a considerable fraction of candidate items that are irrelevant (i.e., not appeared) to any historical behaviors in Amazon datasets; Second, as shown in Figure~\ref{fig:data} (b), more than 50\% of user behavior occurred 30 days ago; Third, in the commercial dataset, there is a large amount of fine-grained user-specific and context information. In contrast, the user's sequential behavior is slightly insufficient. These observations once again prove that the problems we raised before do exist in large-scale recommender systems.
\begin{table*}
    \centering
    \caption{Performance on the public and commercial dataset. The underlined number in each column is the best result among the baseline methods. The percentage in the last row of MIAN is the relative improvement compared to the best baseline. } 
    \vspace{-0.1cm}
    \label{tab:result} 
    \setlength{\tabcolsep}{2mm}{
    \begin{tabular}{p{2.32cm}|p{1.32cm}<{\centering}p{1.32cm}<{\centering}|p{1.32cm}<{\centering}p{1.32cm}<{\centering}|p{1.32cm}<{\centering}p{1.32cm}<{\centering}}
            \toprule
            \multirow{2}{*}{Baseline Methods} &\multicolumn{2}{c}{Books.} &\multicolumn{2}{c}{Electronics.} &\multicolumn{2}{c}{Commercial.}\\ 
            \specialrule{0em}{2pt}{0pt}
            \cline{2-7}\specialrule{0em}{3pt}{0pt}&AUC&Logloss&AUC&Logloss&AUC&Logloss\\
            \midrule
            \specialrule{0em}{4pt}{0pt}
	   (1) LR & 0.7876&0.4401& 0.8214&0.2116 & 0.7584&0.3334\\
            \specialrule{0em}{3pt}{0pt}
	   (2) Wide\&Deep & 0.7932&0.4228& 0.8385& 0.2033 & 0.7611&0.3299\\
            \specialrule{0em}{3pt}{0pt}
	   (3) Deep\&Cross & 0.7995&0.4236& 0.8540& 0.1824 & \underline{0.7646}&0.3197\\
            \specialrule{0em}{3pt}{0pt}
	   (4) DeepFM & 0.8067&0.4188& 0.8636&0.1708 & 0.7609&0.3297\\
            \specialrule{0em}{3pt}{0pt}
	   (5) xDeepFM & 0.8089&0.4174& 0.8683 & 0.1662 & 0.7610&0.3301\\
            \specialrule{0em}{3pt}{0pt}
	   (6) DIN & 0.8145&0.4113& 0.8807&0.1273 & 0.7612&0.3205\\
            \specialrule{0em}{3pt}{0pt}
	   (7) DIEN & \underline{0.8159}&\underline{0.4100}& \underline{0.8836}& \underline{0.1225} & 0.7634& \underline{0.3151}\\
            \specialrule{0em}{3pt}{0pt}
             \cline{1-7}
             \specialrule{0em}{3pt}{0pt}
	    \ \ \textbf{MIAN} & \textbf{0.8209 (+0.61\%)} & \textbf{0.4018 (-2.0\%)} & \textbf{0.8913 (+0.87\%)}&\textbf{0.1136 (-7.3\%)} & \textbf{0.7674 (+0.37\%)}&\textbf{0.3097 (-1.7\%)}\\
            \bottomrule
    \end{tabular}
    }
\end{table*}

\subsection{Baseline Methods}
To evaluate the performance of MIAN, we select the following state-of-the-art approaches which are widely used CTR prediction. 
\begin{itemize}
	\item \emph {\textbf{LR}} \cite{rendle2010factorization}:  It's a widely used baseline and applies linear transformation to model the relationship of all the features.
	\vspace{0.05cm}
	\item \emph {\textbf{Wide\&Deep}} \cite{cheng2016wide}: It jointly trains a linear model and a deep MLP model to the CTR prediction.
	\item \emph {\textbf{Deep\&Cross}} \cite{wang2017deep}: DCN is proposed to handle a set of sparse and dense features, and learn cross high-order features jointly with traditional deep MLP.
	\vspace{0.05cm}
	\item \emph {\textbf{DeepFM}} \cite{guo2017deepfm}: It combines the explicit high-order interaction module with deep MLP module and traditional FM module, and requires no manual feature engineering. 
	\vspace{0.05cm}
	\item \emph {\textbf{xDeepFM}} \cite{lian2018xdeepfm}: It uses Compressed Interaction Network to enumerate and compress all feature interactions, for modeling an explicit order of interactions.
	\vspace{0.05cm}
	\item \emph {\textbf{DIN}} \cite{zhou2018deep}: It's an early work exploits users' historical behaviors and uses the attention mechanism to activate user behaviors in which the user be interested in different items.
	\vspace{0.05cm}
	\item \emph {\textbf{DIEN}} \cite{zhou2019deep}: It is a recent research on CTR modelling with sequential user behavior data. It integrates GRUs with candidate-centric attention for capturing the involved interests.
\end{itemize}
Most of the baseline methods are implemented in Pytorch, based on the public code of DeepCTR \footnote{https://github.com/shenweichen/DeepCTR-Torch} , following their parameter settings (learning rate, batch size, etc).

\subsection{Experiment Setups} 
For MIAN model, we use Adam~\cite{kingma2014adam} as the optimizer, with 0.001 of learning rate, 0.9 of $\beta_1$, and 0.999 of $\beta_2$. The parameters are initialized with truncated normal with a standard deviation of 0.02. Besides, the length of the sequential behavior is set to 30, the dimension of the MLP layer are 32 and 8. The head number of multi-head attention in Transformer and batch size are set to 8 and 128. To avoid overfitting and stabilize the training, a dropout rate of 0.2 and $l$2 regularization with coefficient of 0.001 is used. The models are trained on a machine with 4 NVIDIA Tesla K80 GPUs. Besides, we use two metrics (i.e., AUC and Logloss) for the model evaluation, which are widely used in the CTR prediction.

\subsection{Offline Experimental Results}
Table~\ref{tab:result} shows the offline experimental results of all the baselines and our MIAN on the Amazon datasets and commercial dataset. From the results, we can get several observations.
 
\vspace{0.03cm}
{{(1).}} On the two Amazon datasets, DIN and DIEN have the best performance over all the metrics (i.e., {AUC} and {Logloss}) than the other baselines. The reason that the sequential models outperform xDeepFM and other baseline methods significantly is that the non-sequential models ignore the interactions between user's historical behaviors and the candidate item. It indicate that modeling the sequential behavior is quite important for the representation of user preference and CTR prediction.
 
\vspace{0.03cm}
{{(2).}} We can also observe that MIAN always has the best performance on all three datasets over all the metrics. Compared to the best baseline method, the average relative improvement of MIAN on AUC is 0.62\% and the number is 3.7\% on Logloss, which is significant for the CTR prediction tasks \cite{cheng2016wide,guo2017deepfm}. On the Amazon datasets, MIAN outperforms the best baseline DIEN because our method can efficiently learn the fine-grained multiple interactions of historical behavior, user-specific and contextual interactions, while DIEN only focuses on the user's sequential behavior interactions. To be more specific, MIAN can further capture the internal dependency of different subspaces among the historical behavior through the Pre-LN Transformer, and leverage fine-grained user-specific and contextual interactions efficiently by attention mechanism. 
 
\vspace{0.03cm}
{{(3).}} On commercial dataset which has limited sequential behaviors but plenty of other fine-grained information as shown in Figure~\ref{fig:data} (c), the performance of DCN is slightly better than DIEN by considering multiple feature interactions. However, our method can still beat the best baseline DCN, and achieves improvement up to  3.7\% of AUC, which is a great gain in real industrial scenarios and large scale distributed systems. Besides the sufficient leverage of the high-order interactions as DCN does, we believe that the MIL plays an important role to capture user's sequential behaviors and other fine-grained feature interactions simultaneously.

\vspace{0.03cm}
{{(4).}} Overall, the power of the MIAN lies in three aspects: 1) the effective exploitation of sequential behavior interactions through Transformer, 2) the sufficient utilization of the fine-grained user-specific and context feature interactions through attention mechanism, and 3) the global weighting and integration on the original representations and the learned high-level interactive representations through the global interaction layer. All these three components constitute the MIL, which is the core of our model.

\subsection{Model Efficiency}
DIN and DIEN papers also reported experiment results evaluated on Amazon datasets, and we can observe the same trend of performance that LR $\textless$ Wide\&Deep $\textless$ DeepFM $\textless$ DIN $\textless$ DIEN. The reported AUC and loss are not exactly the same with our paper since some settings are different. For example, DIEN paper processed the data to keep only one record for each user, thus the dataset is no longer the same and the numbers are not directly comparable.

Moreover, the computation time of MIAN is comparable to the most popular sequential models (i.e., DIN and DIEN). Specifically, DIN costs 3,143 seconds, MIAN 3,700s and DIEN 4,073s per epoch on Amazon-book dataset with 4 NVIDIA Tesla K80 GPUs. Compared to DIN, MIAN is a little slower because it incorporates the Pre-LN Transformer unit in IBIM and additional attention layer in IUIM and ICIM. In the other side, MIAN is still faster than DIEN because the GRU computation is more time consuming and difficult to be parallelized during the training process, especially in large scale distributed system. Finally, apart from the Embedding layer, the parameters of MIAN are also acceptable according to \cite{vaswani2017attention}.

\begin{figure}[H]
	\centering
	\includegraphics[width=0.98\columnwidth]{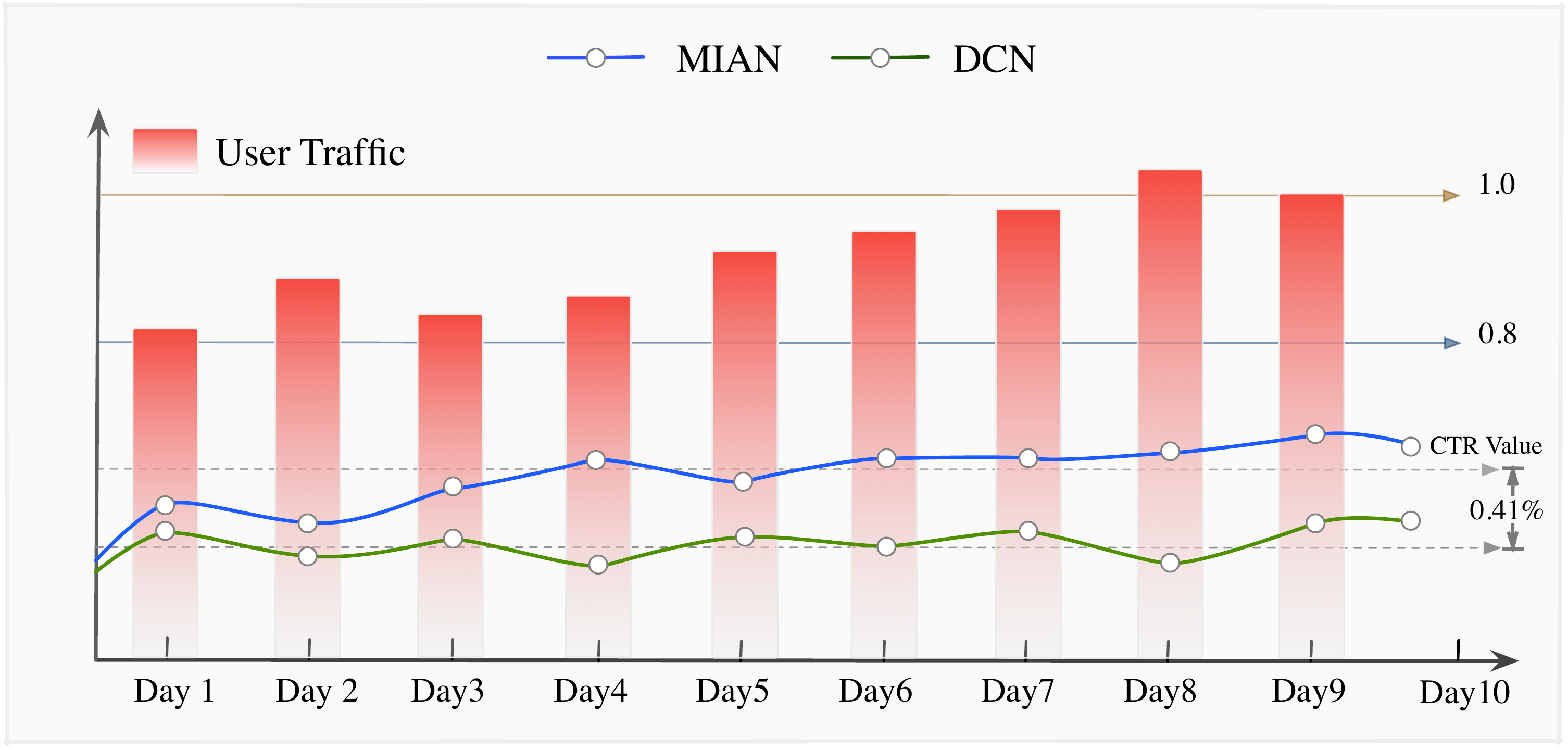}
	\caption{Online A/B testing results of MIAN and DCN.
	}
	\label{fig:online}
\end{figure}

\subsection{Online A/B Testing Results}
To further illustrate the effectiveness of our proposed model, we conducted an online A/B testing in the production environment of Alipay for 10 days. The recommendation system serves at the scale of tens of millions of users in real traffic, hence the traffic is very expensive in the business view. Considering this fact, we only choose the best method DCN in the current online system as the comparison method. 
The candidate items recommended to users include cash reward, coupons, prizes and member credits. The goal is to increase the CTR of the candidate items while constraining the total cost due to limited budget.

Our proposed MIAN model brings a 0.41$\%$ gain in CTR while a 0.27$\%$ drop in cost compared to the best baseline method DCN in a statistically significant level which contributes a considerable business revenue growth. Note that, on a large scale commercial platform, an increase like 0.1\% in CTR value can bring huge benefits. Besides, as shown in Figure~\ref{fig:online}, the CTR gain of our model is consistent during the 10-days online experiment, which demonstrates the effectiveness and stability of our model.

\subsection{Ablation Study and Visualization Study }
\subsubsection{Effectiveness of Multi-Interactive Layer}
\vspace{0.05cm}
To examine the effectiveness of each component in our model, in this subsection, we present multiple ablation studies on the public datasets. The ablation results are summarized in Table~\ref{tab:public_ablation}. Note that removing IBIM means we discard interactions between sequential behavior and the candidate item, use only the original feature from Embedding Layer, and the same comes for removing IUIM and ICIM. Generally, we can observe that each module contributes in a certain degree to the improvement of CTR performance, which demonstrates the reasonability of our proposed architecture.

\begin{table}
    \small
    \centering
    \caption{Ablation performance on public datasets.} 
    \vspace{-0.15cm}
    \label{tab:public_ablation} 
    \setlength{\tabcolsep}{1.4mm}{
    \begin{tabular}{p{3.2cm}p{0.7cm}<{\centering}p{1.0cm}<{\centering}p{1.0cm}<{\centering}p{0.8cm}<{\centering}}
            \toprule
            \multirow{2}{*}{Methods} &\multicolumn{2}{c}{Books.} &\multicolumn{2}{c}{\ \ \ Electronics.}\\
            \specialrule{0em}{2pt}{0pt}
            \cline{2-5}\specialrule{0em}{3pt}{0pt}&AUC&Logloss&AUC&Logloss\\
            \midrule
            \specialrule{0em}{4pt}{0pt}
            1) \textbf{MIAN} &\textbf{0.8209}&\textbf{0.4018}&\textbf{0.8913}&\textbf{0.1136}\\
            \specialrule{0em}{3pt}{0pt}
            2) Ablation w/o IUIM &0.8187&0.4066&0.8875&0.1197\\
            \specialrule{0em}{3pt}{0pt}
            3)  Ablation w/o ICIM &0.8194&0.4058&0.8902&0.1141\\
            \specialrule{0em}{3pt}{0pt}
            4)  Ablation w/o IBIM &0.8154&0.4104&0.8843&0.1214\\
            \specialrule{0em}{3pt}{0pt}
            5)  Ablation w/o GIM &0.8189&0.4060&0.8872&0.1186\\
            \specialrule{0em}{3pt}{0pt}
            6) Ablation w/o IUIM\&ICIM &0.8166&0.4079&0.8846&0.1220\\
            \specialrule{0em}{3pt}{0pt}
            \cline{1-5}
            \specialrule{0em}{3pt}{0pt}
            7) Best Baseline (DIEN) &0.8159&0.4100&0.8836&0.1225\\
            \bottomrule
    \end{tabular}
    }
    \vspace{-0.1cm}
\end{table} 

Taking a close look at the results, removing IBIM hurts the performance of MIAN most among all the settings, which demonstrates that MIAN can effectively leverage the sequential behavior to extract user's historical perference. The next harmful setting is removing both IUIM and ICIM which entirely gives up the user-specific and context features. In a straightforward view, it illustrates the importance of user-specific and context features which are well exploited in our model. Moreover, it is interesting to notice that the performance of MIAN is slightly better than the best baseline DIEN even in this extreme setting. It is a clear indication that IBIM module can sufficiently and effectively exploit the features of user's historical behavior compare with previous sequential methods. In addition, we can observe that removing GIM does hurt the performance of MIAN to a certain degree. It shows that the high-order interactions and the balance of multiple modules are necessary in the whole model for the final CTR prediction.

\subsubsection{Effectiveness of  Pre-LN Transformer}
\begin{figure}[H]
	\centering
	\includegraphics[width=0.9\columnwidth]{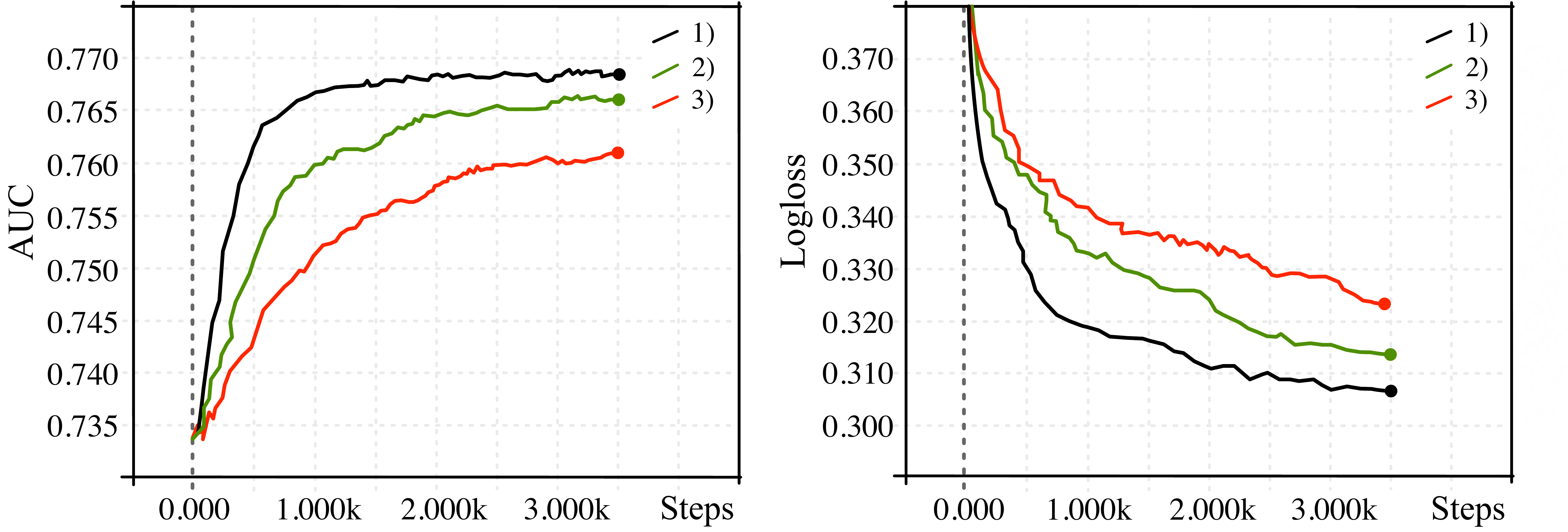}
	\vspace{-0.1cm}
	\caption{Loss convergence of models and the performance comparison on the commercial dataset. 
	}
	\label{fig:loss}
\end{figure}

Table~\ref{tab:pre_ln} summarizes the effects of different user sequential behavior extraction models. We conduct experiment without Transformer or any RNN related layer as DIN does, so that the interactions only occur between the candidate item information and the whole feature representations. We also compare the performance between the original Transformer and Pre-LN Transformer to show the effectiveness of our modification in the real scenario, i.e., commercial recommendation system. Obviously, it hurts the performance without any Transformer since it ignores the dependency among the sequences of user behavior. With the original transformer, the performance increases, which indicates the importance of extracting the associations among user's historical behavior. The Pre-LN transformer further enhances the performance on both metrics, since it makes the training process more stable and speeds up the convergence speed of the mode. The metrics throughout the training process are shown in Figure~\ref{fig:loss}. Corresponding to the Table~\ref{tab:pre_ln}, the method 1) with the black curve is our MIAN model. The method 2) is the variant of MIAN. The method 3) is a simple MIAN, which remove the Pre-LN Transformer directly. 

\begin{figure}[H]
	\centering
	\includegraphics[width=1\columnwidth]{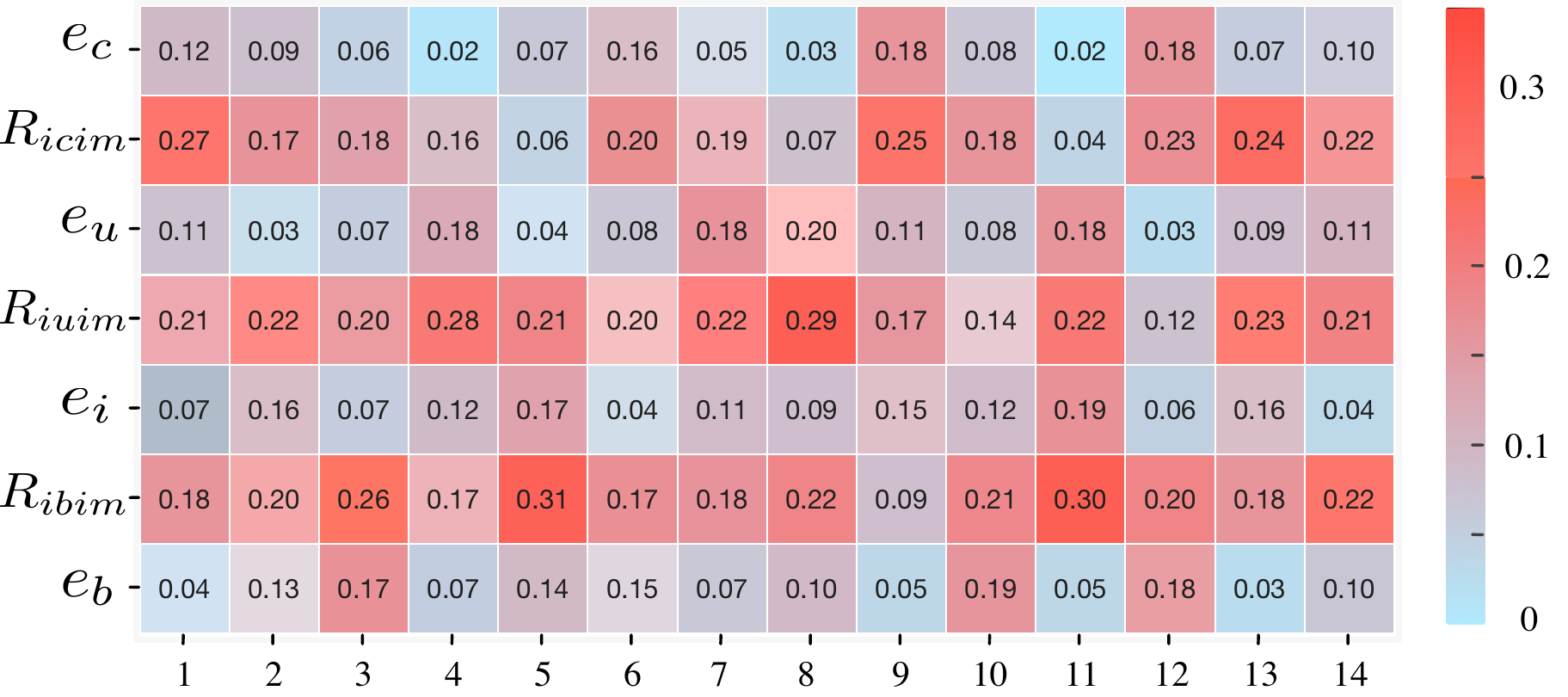}
	\vspace{-0.42cm}
	\caption{Feature weights visualization.}
	\label{fig:attention}
	\vspace{-0.2cm}
\end{figure}

\subsubsection{Visualization of Global-attention}
To intuitively show the importance of fine-grained interactive features and the interpretability of GIM, we randomly select 14 cases from the Amazon dataset and visualize the attention weights in the global attention module presented in Equation~\ref{eq:sixteen} and Figure~\ref{fig:model} (d). As Figure~\ref{fig:attention} shows, among most of the cases presented, the attention weights of interactive feature representations ``$R_{ibim}$'',``$R_{iuim}$'' and ``$R_{icim}$'' are much larger than the original feature representations, i.e., $e_i$, $e_b$, $e_u$, $e_c$. We can also observe several red blocks in the original feature rows, which suggests a possible high-order interaction with this feature. In summary, the visualization results not only demonstrate the importance of the fine-grained feature learning but also indicate that MIAN is able to learn deeply interactive associations between candidate items and multiple fine-grained information. Such findings are quite meaningful for feature design, model understanding and bad case debug in the production scenario. 

\begin{table}
    \small
    \centering
    \caption{Ablation performance on the commercial dataset.} 
    \vspace{-0.15cm}   
     \label{tab:pre_ln}
    \setlength{\tabcolsep}{1.3mm}{
    \begin{tabular}{p{4cm}p{1.7cm}<{\centering}p{1.5cm}<{\centering}}
            \toprule
            \specialrule{0em}{2pt}{0pt}
            \multirow{2}{*}{Methods} &\multicolumn{2}{c}{Commercial dataset.}\\ 
            \specialrule{0em}{2pt}{0pt}
            \cline{2-3}\specialrule{0em}{3pt}{0pt}&AUC&Logloss\\
            \midrule
            \specialrule{0em}{4pt}{0pt}
            1) \textbf{MIAN} (w/Pre-LN Transformer) &\textbf{0.7674}&\textbf{0.3097}\\
            \specialrule{0em}{3pt}{0pt}
            2)  Ablation w/Transformer \footnote{} &0.7662&0.3103\\
            \specialrule{0em}{3pt}{0pt}
            3) Ablation w/o both \footnote{} &0.7612&0.3205\\
            \specialrule{0em}{3pt}{0pt}
            \bottomrule
    \end{tabular}
    
    }
\end{table}
\footnotetext[3]{variant of MIAN, which replace Pre-LN Transformer with transformer.}
\footnotetext[4]{without transformer and Pre-LN Transformer.}

\section{Conclusion} 
In this paper, we presented a novel MIAN to model the fine-grained interactions among items, user sequential behavior, user-specific and context information to further improve the performance of CTR prediction. Specifically, we first designed a Multi-Interactive layer to effectively learn the fine-grained interactions with a Pre-LN Transformer and multiple local interactive modules. Then,  a global interaction module is designed to further capture the high-order interaction between the original feature and the learned interactive representation, and balance the different modules in a flexible way. Experimental results on two public datasets as well as a commercial dataset demonstrate the effectiveness of our proposed approach. The ablation studies further illustrate the effectiveness and explainability of each module, which may bring insights for future work. 

\textbf{Acknowledgements:} This research was partially supported by grants from the National Natural Science Foundation of China (No.s U1605251, 61727809 and 61922073). This research was also supported by Ant Group through Ant Research Program. We special thanks to all of the first-line healthcare providers, physicians and nurses that are fighting the war of COVID-19 against time.

\bibliography{acmart.bib} 
\balance

\end{document}